\documentclass{emulateapj}

\slugcomment{ApJ in press, Feb 3, 2009}

\shorttitle{The Space Density of Compton Thick AGN}
\shortauthors{Treister et al.}

\begin{document}


\title{The Space Density of Compton Thick AGN and the X-ray Background}


\author{E. Treister\altaffilmark{1,2,3}, C. Megan Urry\altaffilmark{4,5,6} and Shanil Virani\altaffilmark{5,6}}


\altaffiltext{1}{Institute for Astronomy, 2680 Woodlawn Drive, University of Hawaii, Honolulu, HI 96822; treister@ifa.hawaii.edu}
\altaffiltext{2}{European Southern Observatory, Casilla 19001, Santiago 19, Chile.}
\altaffiltext{3}{Chandra Fellow}
\altaffiltext{4}{Department of Physics, Yale University, P.O. Box 208121, New Haven, CT 06520.}
\altaffiltext{5}{Yale Center for Astronomy and Astrophysics, P.O. Box 208121, New Haven, CT 06520.}
\altaffiltext{6}{Department of Astronomy, Yale University, PO Box 208101, New Haven, CT 06520.}


\begin{abstract}
We constrain the number density and evolution of Compton-thick Active
Galactic Nuclei (AGN). In the local Universe we use the wide area
surveys from the Swift and INTEGRAL satellites, while for high
redshifts we explore candidate selections based on a combination of
X-ray and mid-IR parameters. We find a significantly lower space
density of Compton-thick AGN in the local Universe than expected from
published AGN population synthesis models to explain the X-ray
background. This can be explained by the numerous degeneracies in the
parameters of those models; we use the high-energy surveys described
here to remove those degeneracies. We show that only direct
observations of CT AGN can currently constrain the number of
heavily-obscured supermassive black holes. At high redshift, the
inclusion of IR-selected Compton-thick AGN candidates leads to a much
higher space density, implying (a) a different (steeper) evolution for
these sources compared to less-obscured AGN, (b) that the IR selection
includes a large number of interlopers, and/or (c) that there is a
large number of reflection-dominated AGN missed in the INTEGRAL and
Swift observations. The contribution of CT AGN to the X-ray
background is small, $\sim$9\%, with a comparable contribution to the
total cosmic accretion, unless reflection-dominated CT AGN significantly 
outnumber transmission-dominated CT AGN, in which case their contribution
can be much higher. Using estimates derived here for the accretion luminosity over 
cosmic time we estimate the local mass density in supermassive black holes and 
find a good agreement with available constraints for an accretion efficiency of $\sim$10\%.
Transmission-dominated CT AGN contribute only $\sim$8\% to total black hole growth.
\end{abstract}


\section{Introduction}

It is now clear that most accretion of mass onto supermassive black
holes is obscured from our view (e.g.,
\citealp{fabian99b,treister04}). Observations of nearby Active Galactic Nuclei
(AGN) suggested that the local ratio of obscured to unobscured sources
is $\sim$4:1 (e.g. \citealp{risaliti99}). A similarly high fraction of
obscured AGN has been used to explain the spectrum and normalization
of the extragalactic X-ray Background (XRB), as shown by the latest
AGN population synthesis models \citep{treister05b,gilli07}. The XRB
gives an integral constraint to the AGN population and its evolution;
the most recent deep surveys show that $\sim$90\% of the observed 2--8
keV XRB radiation can be attributed to resolved AGN
(\citealp{hickox06}, and references therein).

The most obscured AGN known are those in which the neutral hydrogen
column density ($N_H$) in the line of sight is higher than the inverse
Thomson cross section,
$N_H$$\simeq$1.5$\times$10$^{24}$~cm$^{-2}$. These are the so-called
Compton-thick (CT) AGN. If the obscuring column density is smaller
than $\sim$10$^{25}$~cm$^{-2}$, direct emission from the nucleus is
still visible at energies greater than $\sim$10~keV, while the
radiation at lower energies is completely obscured by
photoelectric absorption; in this case we have a
transmission-dominated CT AGN. For sources with
$N_H$$>$10$^{25}$~cm$^{-2}$ the X-ray emission is significantly
affected by Compton scattering at all energies, fully obscuring the
direct AGN emission and leaving only the much fainter reflection
component to be detected; these are reflection-dominated AGN.

Contrary to the situation for less obscured sources, not much is known
about the number density of CT AGN. Thanks to the deep Chandra and
XMM-Newton surveys it is now clear that the fraction of
moderately obscured, Compton-thin, AGN is on average $\sim$3/4 of all AGN, and
is higher at lower luminosities
\citep{ueda03,treister05b,barger05} and higher redshifts
\citep{lafranca05,ballantyne06,treister06b}, but there are no
comparable constraints on the number of CT AGN. About a dozen CT AGN
have been identified in the local Universe (\citealp{dellaceca08b} and
references therein). In fact, two of the three nearest AGN are Compton
thick (NGC4945 and the Circinus Galaxy; \citealp{matt00}). Based on a
sample of 49 local Seyfert 2 galaxies, \citet{guainazzi05} estimated
that $\sim$50\% of all obscured AGN ($N_H$$>$10$^{22}$~cm$^{-2}$) are
Compton thick, and similar estimates were made by \citet{risaliti99}
based on much smaller numbers.

However, so far there has been no systematic study of the statistical
properties of CT AGN with a well-defined selection function. Hence,
while it has been hypothesized that CT AGN can contribute up to
$\sim$30\% of the XRB \citep{gilli07} and represent a significant
fraction of the cosmic accretion onto supermassive black holes
\citep{marconi04}, this has not been demonstrated. Now, thanks to the
wide-area surveys at high energies performed with the INTEGRAL/IBIS
\citep{beckmann06,krivonos07} and the Swift/Burst Alert Telescope (BAT;
\citealp{tueller08}), it is possible to study a well-defined sample of
CT AGN in the local Universe. Furthermore, since most of the absorbed
energy is re-emitted at mid-IR wavelengths, deep observations with the
Spitzer observatory can be used to select CT AGN candidates at high
redshift, $z$$\sim$2 \citep{daddi07,fiore08,alexander08}, yielding an upper limit to
the number density of these sources.

In this paper we constrain the number density of CT AGN in the local
Universe from high-energy observations, and at high redshift using a
combination of X-ray and mid-IR data. We compare the observed numbers
of CT AGN with expectations from AGN population synthesis models that
explain the XRB emission and we study the degeneracies affecting these
models. Finally, we compute the implied density of supermassive black holes as
a function of redshift, including transmission-dominated CT accretion.
When required, we assume a $\Lambda$CDM cosmology with $h_0$=0.7,
$\Omega_m$=0.3 and $\Omega_\Lambda$=0.7, in agreement with the most
recent cosmological observations \citep{spergel07}.

\section{The Local Sample of CT AGN}

One of the best ways to find CT AGN is by observing at high energies,
namely E$>$10~keV. The hard X-ray spectrum of a CT AGN is characterized by
at least three components: an absorbed power law with an upper cutoff
at $\sim$300~keV (e.g., \citealp{nandra94}), a Compton reflection hump
which peaks at $\sim$30~keV \citep{magdziarz95} and an iron K$\alpha$
line at $\sim$6.4~keV. Not all components are clearly observed in all
AGN (e.g., \citealp{soldi05,beckmann04}), perhaps because of the low
signal-to-noise of some of the observations. One clear advantage of
high-energy observations is that photoelectric absorption has minimal
effects, so transmission-dominated CT AGN can be easily detected. It
is only when the source becomes reflection-dominated that the emission
at E$>$10~keV is affected.

Current observations at E$>$10~keV are available only at relatively
high fluxes, and hence low redshifts, $z$$<$0.05. While BeppoSAX
\citep{boella97} was successfully used for targeted observations of
known Seyfert galaxies, it is only now thanks to the International
Gamma-Ray Astrophysics Laboratory (INTEGRAL; \citealp{winkler03}) and
Swift \citep{gehrels04} satellites that large-area surveys at these
energies have been done.

Using the IBIS coded-mask telescope \citep{ubertini03}, INTEGRAL
surveyed $\sim$80\% of the sky down to a flux of 5 mCrab in the
17-60~keV. The catalog of \citet{krivonos07} reports the properties of
130 sources detected in these all-sky observations and classified as
AGN. A large number of unidentified sources remain in this catalog,
48, but only seven are found a high galactic latitude
($|$$b$$|$$>$5$^o$), and thus of likely extragalactic origin. Five of
the 130 AGN are CT AGN.

We carried out a very deep survey with INTEGRAL/IBIS, with a total
exposure time of $\sim$3 Msec, in the XMM-Large Scale Survey (XMM-LSS)
region, reaching a flux limit of $\sim$3$\times$10$^{-12}$~erg~cm$^{-2}$s$^{-1}$ in the 20-40 keV band
(S. Virani in prep.). A total of 15 sources, all AGN, are found in
this survey, including the prototypical CT AGN, NGC 1068. We also
found another CT AGN candidate, not detected in X-rays before. However,
an accurate $N_H$ determination has not been obtained for this source yet, and
hence it is not included in our sample. NGC 1068 was also included in the catalog
of \citet{krivonos07} and hence is already in our sample.

Recently, \citet{tueller08} presented a catalog of 103 AGN detected in
an all-sky survey with the Swift/BAT telescope. 
The 14 sources classified as blazars and BL Lac 
were excluded from our sample. 89 of the remaining AGN are
at high galactic latitudes, $|$$b$$|$$>$15$^o$, where only one
source remains unidentified. The fraction of unidentified sources is
much smaller for Swift compared to INTEGRAL because of follow-up
observations with Swift's dedicated X-ray telescope. In the
\citet{tueller08} catalog there are five AGN with estimated
$N_H$ greater than 10$^{24}$~cm$^{-2}$. However, we caution that these
$N_H$ measurements were obtained by fitting a single absorbed power-law to
the X-ray spectrum, while it is known that heavily absorbed AGN have
more complex spectra (e.g., \citealp{vignati99,levenson06}).

We added to our sample the source NGC 7582, which has
$N_H$$\sim$10$^{23}$~cm$^{-2}$ in \citet{tueller08} but has been shown
with XMM/Newton data to have a very complex spectrum with strong
evidence for mildly Compton-thick absorption,
$N_H$$\sim$10$^{24}$~cm$^{-2}$ \citep{piconcelli07}. With the improved
sensitivity of the Suzaku telescope \citep{mitsuda07}, it is possible
to perform detailed X-ray spectroscopy for some of the sources
detected by Swift/BAT included in the catalog of \citet{tueller08},
revealing in some cases Compton-thick absorption. We added to our
sample the source NGC 5728, which as reported by \citet{comastri07}
from Sukazu observations is obscured by a Compton-thick gas with
$N_H$$\simeq$2.1$\times$10$^{24}$~cm$^{-2}$. We also added the source
ESO 005--G004 which based on the Suzaku observations reported by
\citet{ueda07} is a heavily obscured, CT AGN.

In summary, INTEGRAL and Swift found 130 and 103 AGN, respectively, in their wide area 
surveys; 76 sources ($\sim$58\%) were detected by both surveys. (This fraction is not larger due to the
differences in sky coverage and the non-uniform depth of the
observations.) We then found 15 AGN in deep 3 Msec INTEGRAL
observations, one of them the CT AGN NGC1068. INTEGRAL detected five CT AGN, while Swift found
eight. However, there is incomplete overlap between the two samples
and we note that the disparate energy ranges make direct comparison of
fluxes difficult. The INTEGRAL-detected CT sources are: NGC4945,
Circinus, Markarian 3, NGC3281 and NGC1068; the Swift/BAT CT AGN are:
NGC4945, Markarian 3, NGC3281, NGC7582, NGC5728, NGC5252, NGC6240 and
ESO 005--G004.

The number of CT AGN found by these surveys is surprisingly low,
compared to the sample of known CT AGN in the local Universe. In a
study of optically-selected Seyfert 2 galaxies with hard X-ray
information, \citet{risaliti99} found 16 CT AGN in a total of 45
Seyfert galaxies, although four were later demonstrated to most likely
not be Compton thick (NGC 1386, IC 3639, NGC 5005 and NGC 4939;
\citealp{maiolino98,ghosh07,gallo06}). Of the remaining 12 CT
sources, three were detected by Chandra and/or XMM, while the
rest are mostly reflection-dominated sources, too faint to be
detected by either INTEGRAL or Swift even though they are nearby,
moderate luminosity AGN. In fact, \citet{awaki08} recently confirmed
the CT nature of NGC 2273, one of the sources in the
\citet{risaliti99} sample, which is however too faint to be detected
by INTEGRAL or Swift.

Recently, \citet{dellaceca08b} published a list of 18 CT AGN with
detections at E$>$10~keV. Of these, seven were detected by Swift and/or
INTEGRAL, while the remaining 11 were studied with pointed BeppoSAX
observations, and are typically fainter than the INTEGRAL/Swift
detection threshold. We use all these samples, suitably amended as necessary,
to place constraints on the number density of CT AGN.

An alternative way to find CT sources is by studying the water maser emission in
AGN. Because large amounts of molecular gas are required to produce the maser
amplification, AGN with detected water maser emission are in general heavily 
obscured along the line of sight. In fact, \citet{greenhill08} recently reported that from 
a sample of 42 AGN known to show water maser emission, 95\% have 
$N_H$$>$10$^{23}$~cm$^{-2}$ and 60\% are Compton-thick. Since these
AGN were not detected at high energies by either INTEGRAL or Swift we
do not include them here; however, we note that water maser emission appears to be a highly efficient
way to identify a large number of heavily obscured sources.

\subsection{The Log N-log S Distribution}

\begin{figure}
\begin{center}
\plotone{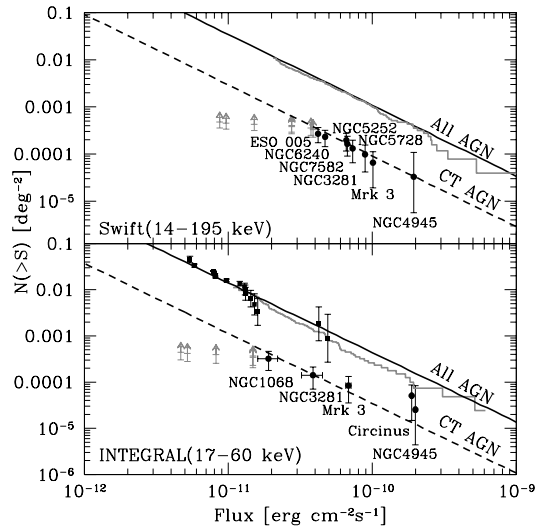}
\end{center}
\caption{LogN-$\log$S distribution for AGN detected at high energies. 
The {\it gray line} in the {\it top panel} shows the AGN in the 
well-defined Swift/BAT samples 
in the 14-195 keV band \citep{tueller08}, while the {\it
bottom panel} shows the INTEGRAL sources \citep{krivonos07} in the
17-60 keV band. {\it Solid squares} show the 15 sources detected in the
deep 3 Msec INTEGRAL observations of the XMM-LSS field (S. Virani in prep.).
{\it Solid circles} mark the CT AGN detected with Swift ({\it top
panel}) and INTEGRAL ({\it bottom panel}). The {\it black solid
lines} show the expected AGN $\log$N-$\log$S from the population
synthesis model of \citet{treister05b}, which at these fluxes corresponds to 
a Euclidean distribution. The {\it dashed lines} mark
the Euclidean slope normalized to the number of Swift and INTEGRAL 
CT AGN. The {\it gray} lower limits show the previously-known 
transmission-dominated 
AGN with hard X-ray observations, not detected in the INTEGRAL or Swift surveys. These are lower 
limits since they were selected from pointed observations and are thus highly incomplete.}
\label{logn_s}
\end{figure}

Figure~\ref{logn_s} shows the cumulative number counts of AGN, with CT
sources shown separately, as a function of hard X-ray flux. In order
to avoid the necessity of specifying a standard spectrum to convert
fluxes to different energy bands, we show the INTEGRAL and Swift
sources separately, but note that a good agreement (within $\sim$40\%)
in the normalization between the two distributions exist if a standard
band conversion is assumed. At these high fluxes the slope of the
$\log$~N-$\log$~S is Euclidean, implying an uniform spatial
distribution, as expected given the low redshifts of these sources. We
also compare with the distribution predicted by the AGN population
synthesis model with which \citet{treister05b} fit the XRB, and find
in general good agreement in slope and normalization.

The $\log$~N-$\log$~S relations for the five CT AGN detected by
INTEGRAL and the eight sources observed by Swift/BAT are also
consistent with Eculidean slopes, with normalizations of
10$^{-4}$~deg$^{-2}$ at fluxes of $\sim$5$\times$10$^{-11}$ and
$\sim$9$\times$10$^{-11}$ erg~cm$^{-2}$s$^{-1}$ in the INTEGRAL and
Swift bands, respectively. 
(For a power law spectrum with $\Gamma=1.8$, for example, 
$F_{14-195, Swift} \sim 2.3 F_{17-60, INTEGRAL}$ 
for most column densities $N_H \lesssim 24$~cm$^{-2}$.)
About twice as many CT AGN at low redshifts
were reported in the sample of \citet{risaliti99}. Since all these CT
AGN have detections at high energies, we plot them in
Fig.~\ref{logn_s}. Clearly, they fall significantly below the
extrapolation of the observed $\log$~N-$\log$~S, suggesting high
levels of incompleteness in the \citet{risaliti99} sample. This is not
surprising given that these sources do not come from a flux-limited
survey but from pointed observations.

A possible source of incompleteness in our sample of CT AGN comes from
the difficulty in measuring the amount of absorption in these sources,
given that they have in general very complex X-ray spectra. As noted
by \citet{tueller08}, sources that are not well-characterized in
X-rays by an absorbed power-law are good candidates to be heavily
obscured AGN. In their Swift/BAT sample, a total of 46 sources with
complex spectra were reported. Of those, 18 have an optical
classification of Seyfert 1.5 or lower, and hence it is very unlikely
that they are CT AGN. Considering the very extreme assumption that the
remaining 28 sources are all CT AGN increases the normalization of the CT
AGN $\log$~N-$\log$~S by only a factor of $\sim$2. This is because a large
fraction of the complex-spectrum sources have fluxes fainter than that
of NGC6240, one of the faintest confirmed CT AGN in the Swift
sample. In any case, it is important to remark that according to a
detailed study by \citet{winter08}, while all these complex-spectrum sources are
highly obscured, only half a dozen have
some evidence of Compton-thick column densities. Hence, we 
conclude that the observed $\log$~N-$\log$~S is not affected
significantly by this possible source of incompleteness.

For the transmission-dominated AGN in our sample (i.e., excluding
NGC1068), we find volume densities for L$_X$$>$10$^{42}$~erg~s$^{-1}$
of 5.5$^{+8.6}_{-3.1}$$\times$10$^{-5}$ Mpc$^{-3}$; and for
L$_X$$>$10$^{43}$~erg~s$^{-1}$ of 2.2$^{+2.9}_{-1.1}$$\times$10$^{-6}$
Mpc$^{-3}$.  Because this is a flux-limited sample, luminosity and
redshift are strongly correlated.  For example, a source with X-ray
luminosity of 10$^{42}$, 10$^{43}$ or 10$^{44}$~erg~s$^{-1}$ can only
be detected up to $z$$\simeq$0.005, 0.015 or 0.045,
respectively. Thus, the source densities inferred here are valid only
up to these limiting redshifts, corresponding to distances of
$\sim$21, 63 and 190 Mpc. Also, because there is a significant
correlation between luminosity and fraction of obscured sources (e.g.,
\citealp{ueda03,barger05}), CT AGN are found only up to $z$=0.024 in
this sample, even though unobscured sources have been found by
Swift/BAT up to $z$$\sim$0.1. That is, CT AGN preferentially have low
luminosities, so they are found mostly at low redshift.  In effect,
the flux limit prevents us from detecting the many CT AGN at higher
redshift. The derived volume densities for CT AGN at $z$$\simeq$0 are
fully consistent with the values derived by \citet{dellaceca08a} from
three INTEGRAL sources only.

Taking into account the densities reported here, the number of CT AGN
relative to the X-ray-selected AGN population is
5.3$\times$10$^{-5}$/2.2$\times$10$^{-4}$=24\% for sources in the
L$_X$=10$^{42-43}$~erg~s$^{-1}$ range, while for sources with
$L_X$=10$^{43-44}$~erg~s$^{-1}$ this fraction is
2.2$\times$10$^{-6}$/2.9$\times$10$^{-5}$=7.5\%. This calculation uses
the luminosity function of \citet{ueda03}. However, similar numbers
are obtained if the \citet{lafranca05} luminosity function is used
instead. Hence, the relative fraction of CT AGN decreases by about a
factor of 3 for an order of magnitude increase in luminosity.  For
Compton-thin sources, according to \citet{treister08b}, the fraction
of obscured sources ($N_H\geq10^{22}$~cm$^{-2}$) decreases from 100\%
at $L_X$=10$^{42}$~erg~s$^{-1}$ to $\sim$60\% at
L$_X$=10$^{43}$~erg~s$^{-1}$, implying a decrease of about a factor of
2.  Therefore, the decrease in the fraction of CT AGN is comparable to
the decrease in the fraction of obscured Compton-thin AGN, a
reasonable agreement given the statistical errors in our sample. This
is in agreement with the conclusions of \citet{fiore09}, who found a
similar decrease in CT AGN with increasing luminosity in their
high-redshift IR-selected sample.

\subsection{N$_H$ Distribution}

\begin{figure}
\begin{center}
\plotone{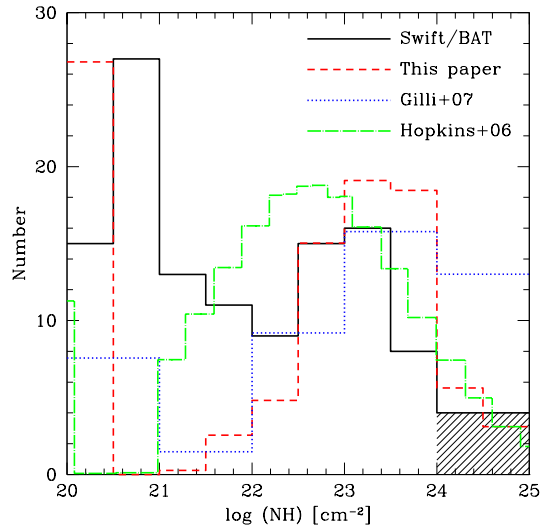}
\end{center}
\caption{Distribution of neutral hydrogen column density ($N_H$) for
the AGN detected in the Swift/BAT survey ({\it solid histogram}). The
distribution is to first approximation flat, but shows a significant decrease in the number of AGN with
 $N_H$$>$10$^{24}$~cm$^{-2}$. The {\it dotted histogram} shows the
$N_H$ distribution used in the \citet{gilli07} AGN population
synthesis model, normalized to the number of sources in the Swift/BAT
survey, while the {\it dot-dashed} line shows the $N_H$ distribution
predicted by the galaxy-merger models of \citet{hopkins06}. The {\it dashed
histogram} shows the $N_H$ distribution assumed by the
\citet{treister05b} model for 10$^{20}$$<$$N_H$$<$10$^{24}$~cm$^{-2}$, 
modified for 
the flux limit of the Swift/BAT survey and normalized to
the same number of sources. The discrepancies at low $N_H$ are not relevant to
the present work (see \S 2.2 for details). For $N_H$$>$10$^{24}$~cm$^{-2}$, the {\it dashed region}
shows the number of CT AGN allowed by the current observations.}
\label{nh_dist}
\end{figure}

A key ingredient in our understanding of the AGN population and of the
properties of the obscuring material is the distribution of
line-of-sight column densities, parametrized in terms of the neutral
hydrogen column density, $N_H$. In Fig.~\ref{nh_dist} we show the
observed $N_H$ distribution for the sources in the Swift/BAT sample of
\citet{tueller08} obtained from very simple spectral fitting assuming
an intrinsic absorbed power-law spectrum; the distribution from the
AGN population synthesis models of \citet{treister05b}, adapted to the
flux limit of the Swift/BAT sample; the distribution assumed by
\citet{gilli07}; and the distribution predicted by the galaxy
evolution models of \citet{hopkins06}.

The $N_H$ distribution observed in the Swift sample is relatively
flat, before a strong decline at $N_H$$>$10$^{24}$~cm$^{-2}$. This
decline corresponds to a relatively low number of CT AGN in this
sample, as mentioned before. In contrast, the $N_H$ distribution used
in the XRB population synthesis model of \citet{treister05b} had
roughly the same number of sources with $N_H$ in the
10$^{23}$-10$^{24}$~cm$^{-2}$ and 10$^{24}$-10$^{25}$~cm$^{-2}$
ranges, because of an incorrect assumption about the normalization of
the HEAO1-A2 X-ray background.  This translates into a discrepancy of
a factor of $\sim$3 more CT AGN assumed by that model than is observed
in the BAT sample. A similar unrealistically high number of CT AGN was
assumed in the work of \citet{gilli07}. The relationship between the
number of CT AGN and the XRB is explored in detail in \S3 below.

For relatively unobscured sources, $N_H$$<$10$^{22}$cm$^{-2}$, the
\citet{tueller08} $N_H$ distribution is significantly different from
that observed in the deepest X-ray fields. For example, for the
sources in the Chandra Deep Field North and South, \citet{treister04}
reported a sharp peak at $N_H$=10$^{20}$~cm$^{-2}$ and almost no AGN
in the 10$^{20}$-10$^{21}$~cm$^{-2}$ range. The discrepancy at low
values of $N_H$ between the Swift/BAT and the deep fields samples
could be due to the difficulty in measuring low $N_H$ values at higher
redshifts, as discussed by e.g., \citet{akylas06}. The $N_H$
distribution in the population synthesis model of \citet{treister05b}
matches well the observed distribution in the Chandra deep fields, but
has some significant differences with the distribution of
\citet{tueller08}. This discrepancy however is not relevant for our
present work, which focuses on the most obscured AGN.

\subsection{Comparison with Models}

The $N_H$ distribution for AGN is predicted by the galaxy evolution
models of \citet{hopkins06}.  They assumed that AGN are fueled solely
by mergers of gas-rich galaxies, and the $N_H$ distribution was
derived by integrating the amount of gas along the line of sight for
the simulated galaxies. The resulting distribution peaks at
$N_H$$\sim$10$^{22.5}$~cm$^{-2}$ and declines strongly towards higher
column densities, in reasonable agreement with the number of CT
sources reported here. It is interesting to note that in the Hopkins
et al. model the obscuring gas is located $\sim$100-200 pc from the
nucleus.  Such a large scale for the obscuration disagrees with
observations of a few nearby AGN using near-IR interferometry, which
show outer radii for the obscuring material of $\sim$3 pc for NGC1068
\citep{jaffe04} and $\sim$2 pc for the Circinus galaxy
\citep{tristram07}, for example. Similarly, the latest torus models
predict small scales for the obscuring material, $\leq$10 pc
\citep{nenkova08b}, although fitting the IR AGN spectra does not
provide a very strong constraint to the torus size. Further
comparisons will provide an interesting test of the Hopkins et
al. model.

The fact that a relatively small number of AGN with
$N_H$$>$10$^{24}$~cm$^{-2}$ is observed can be used to constrain the
nature of the obscuring material. This lack of CT AGN can be
interpreted either in the context of a clumpy torus (e.g.,
\citealp{krolik88,nenkova08a}) or a smooth distribution (e.g.,
\citealp{pier92}). For example, the $N_H$ distribution of
\citet{treister05b}, presented in Fig.~\ref{nh_dist}, assumed a smooth
torus with a single equatorial column density of
10$^{25}$~cm$^{-2}$. To accommodate a smaller number of CT AGN while
still matching the observed $N_H$ distribution for Compton-thin
sources would require a distribution of equatorial densities, in which
only a small fraction of the AGN reach the CT levels for
nearly-equatorial line of sights. In the case of a clumpy torus, the
explanation is perhaps more natural; the small fraction of CT AGN
implies that only a few sources have a large number of clouds, e.g.,
$> 10$ clouds for the models of \citet{nenkova08a}.

\section{CT AGN and the X-ray Background}

\subsection{Parameter Degeneracies}

The spectrum of CT AGN at high energies, even for
transmission-dominated sources, is often dominated by the Compton
reflection component (e.g., \citealp{matt00}), which has a strong peak
at E$\sim$30 keV \citep{magdziarz95}. The observed spectrum of the
XRB, which we now know is just the integrated emission from previously
unresolved AGN, also has a peak at about the same energy
\citep{gruber99}. Hence, it was suspected for a long time that CT AGN
provide a significant contribution to the XRB emission. In fact, in
the early work of \citet{comastri95} the contribution of CT AGN to the
XRB is $\sim$20\%, similar to the value assumed in the population
synthesis models of \citet{ueda03}, \citet{treister05b}, and
\citet{gilli07}; \citet{shankar08a} report a slightly higher
contribution of $\sim$30\% at $\sim$20~keV.

\begin{figure}
\begin{center}
\plotone{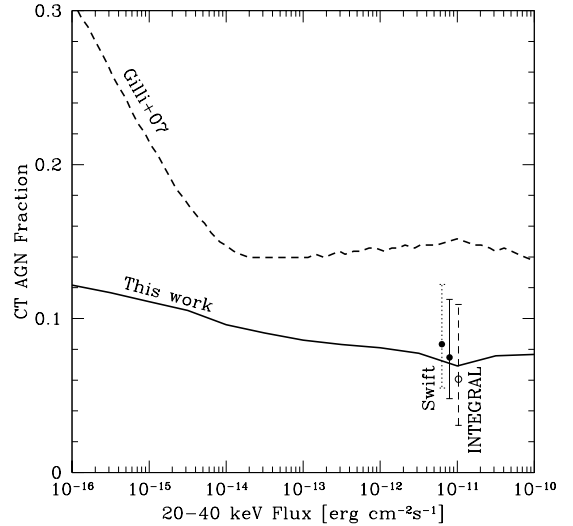}
\end{center}
\caption{Measured fraction of AGN that are Compton-thick in the 
INTEGRAL ({\it open circle}; \citealp{krivonos07}) 
and Swift/BAT ({\it filled circles}; \citealp{tueller08}) samples. The filled
circle with solid error bars shows the fraction using the identified
sources only, while the circle with dotted error bars assumes that the
one unidentified source is a CT AGN. The {\it solid line} shows the
fraction of CT AGN from the modified \citet{treister05b} population synthesis model; the original assumption was a factor of $\sim$4 too high, 
so was modified as described in the present text to match the Swift/BAT and INTEGRAL observations. The {\it dashed line} shows 
the fraction of CT AGN in the model of \citet{gilli07}, which is a factor of $\sim$3 higher than observations, and increases
sharply at faint fluxes because of the assumed steep dependence of the fraction of obscured sources on luminosity.}
\label{frac_ct}
\end{figure}

Because it is very hard to measure the number density of CT AGN, even
locally, AGN population synthesis models have assumed it to be a fixed
fraction of the obscured, Compton-thin sources, typically $\sim$0.5-1
times as many. In Figure~\ref{frac_ct} we show the fraction of all AGN
that are Compton-thick, compared to the observed fraction in the
INTEGRAL and Swift samples as a function of hard X-ray flux. At fluxes
of $\sim$10$^{-11}$~erg~cm$^{-2}$~$^{-1}$ the fraction of CT AGN in
the model of \citet{gilli07} is $\sim$15\%, while the observed value
is $\sim$6$\pm$5\% for INTEGRAL, and $\sim$8$\pm$3\% for the Swift
sample.  For comparison, in Figure~\ref{frac_ct} (solid line) we show
the predicted CT AGN fraction as a function of flux for the model of
\citet{treister05b} with the number of CT AGN modified to match the
INTEGRAL and Swift observations presented here. The
\citet{treister05b} model assumes a nearly constant fraction of CT
AGN, while the \citet{gilli07} model assumes a steep increase in the
number of CT AGN at fluxes fainter than
$\sim$10$^{-14}$~erg~cm$^{-2}$~s$^{-1}$. This is due to the assumed
luminosity dependence of the fraction of obscured sources in the
\citet{gilli07} model, which decreases steeply above luminosities of
$\sim$10$^{43}$~erg~s$^{-1}$ and is flat at lower luminosities. While
such faint fluxes are still out of reach for current hard X-ray
observatories, it will be possible to test this flux regime with
NuSTAR and the International X-ray Observatory (IXO). However, the
luminosity dependence of the fraction of obscured AGN assumed by the
model of \citet{gilli07} can already be ruled out, in particular at
high luminosities, by observations of Compton-thin AGN at lower
energies \citep{hasinger08,treister08b}.
 
The fact that the XRB does not constrain the number density of CT AGN
can be explained by strong degeneracies in other parameters, like the
assumed spread in spectral index \citep{gilli07,shankar08a},
high-energy cutoff, etc. When trying to constrain the number density
of CT AGN the most important parameter is the normalization of the
Compton reflection component, which is directly related to the AGN
luminosity at $\sim$30 keV, where the CT AGN contribution is
maximal. Given that even the nearest AGN have only low signal-to-noise
observations at E$>$10~keV, the normalization of the Compton
reflection component is not well constrained by observations of
individual sources. From a sample of 22 Seyfert galaxies, excluding CT
sources, \citet{malizia03} concluded that both obscured and unobscured
sources have similar reflection components with normalization values
in the range R$\sim$0.6-1 (in units of 2$\pi$).  A similar value of
R$\simeq$1 was reported by \citet{perola02} based on BeppoSAX
observations of a sample of 9 Seyfert 1 galaxies.  Although with large
scatter, normalizations for the average reflection component of 0.9
for Seyfert 1 and 1.5 for Seyfert 2 were measured from BeppoSAX
observations of a sample of 36 sources \citep{deluit03}. Early
population synthesis models for the XRB assumed values of R=1.29 for
unobscured sources and 0.88 for obscured AGN
\citep{comastri95,gilli99}. In contrast, the models of \citet{ueda03},
\citet{treister05b} and \citet{ballantyne06} assumed a constant value
of R=1 (equivalent to a solid angle of 2$\pi$) for both obscured and
unobscured sources. \citet{gilli07} assumed the same normalizations as
\citet{comastri95}, R$\sim$1.3 and 0.88; however, for high-luminosity
sources, $L_X$$>$10$^{44}$~erg~s$^{-1}$ , the reflection component was
not included (R=0).

\begin{figure}
\begin{center}
\plotone{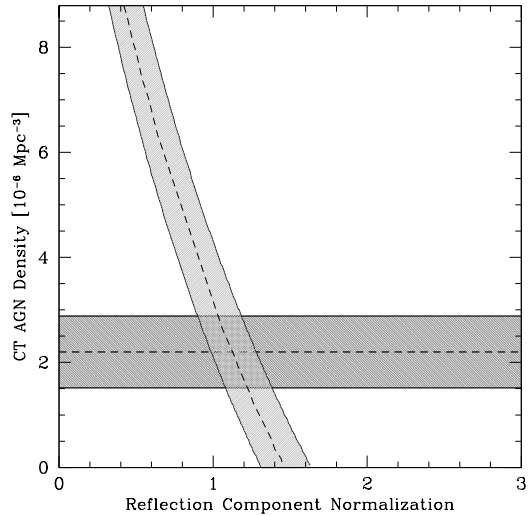}
\end{center}
\caption{Degeneracy of the local density of CT AGN and the normalization of the Compton
reflection component subject to the constraint of either the XRB intensity or the number of CT AGN in hard X-ray surveys. 
The {\it dark gray region} shows the space density obtained from the 10 CT AGN detected by Swift and INTEGRAL in complete samples at 
$z$$\sim$0, including 1-$\sigma$ statistical fluctuations. The {\it light gray region} shows the constraints to these 
parameters given by the intensity of the XRB in the 20-40 keV band, assuming a 5\%
uncertainty in the XRB intensity. The correct values of these parameters must be at the intersection between these
two regions, namely, a normalization of the Compton reflection
component of $\sim$1 and a CT number density of $\sim$2$\times$10$^{-6}$~Mpc$^{-3}$.}
\label{dens_comp}
\end{figure}

For a given number of CT AGN, the resulting intensity of the XRB at
$\sim$30~keV is directly linked to the assumed normalization of the
reflection component.  In Figure~\ref{dens_comp} we show the values of
the normalization of the Compton reflection component and CT AGN
number density that produce XRB intensities in the 20-40~keV region
consistent with the latest observed values from INTEGRAL
\citep{churazov07} and Swift \citep{ajello08}. This is the energy
range in which the contribution of CT AGN to the XRB is maximal and
hence can be best constrained.  For comparison, the parameters assumed
by the model of \citet{gilli07} lie in the upper left region of
Figure~\ref{dens_comp}, at a density of CT AGN roughly 3$\times$
higher than the observed value of $\sim$2$\times$10$^{-6}$~Mpc$^{-3}$
and average Compton reflection component normalization of
$\sim$0.6. (The latter is a rough estimate, since they assumed
$R=0.88$ for obscured sources at low-luminosities, and $R=0$ at high
luminosities.)  The model of \citet{treister05b} assumed a similarly
high number of CT AGN and a higher normalization of the Compton
reflection component, and hence resulted in a higher intensity of the
XRB, consistent with the HEAO-1 \citep{gruber99} measurements
increased by 40\%, similar to what was assumed by \citet{ueda03} and
\citet{ballantyne06} in order to match the observations of the XRB at
lower energies by Chandra and XMM.  Such a high value of the XRB
intensity at E$\sim$10-50~keV is now ruled out by new INTEGRAL
\citep{churazov07}, Swift \citep{ajello08} and BeppoSAX
\citep{frontera07} data. Given the degeneracies with other model
parameters, it is unlikely that the XRB could be used to constrain the
average value of R. High signal-to-noise observations of individual
sources at E$>$10~keV are required for this purpose.

\subsection{A New X-ray Background Fit}

\begin{figure}
\begin{center}
\plotone{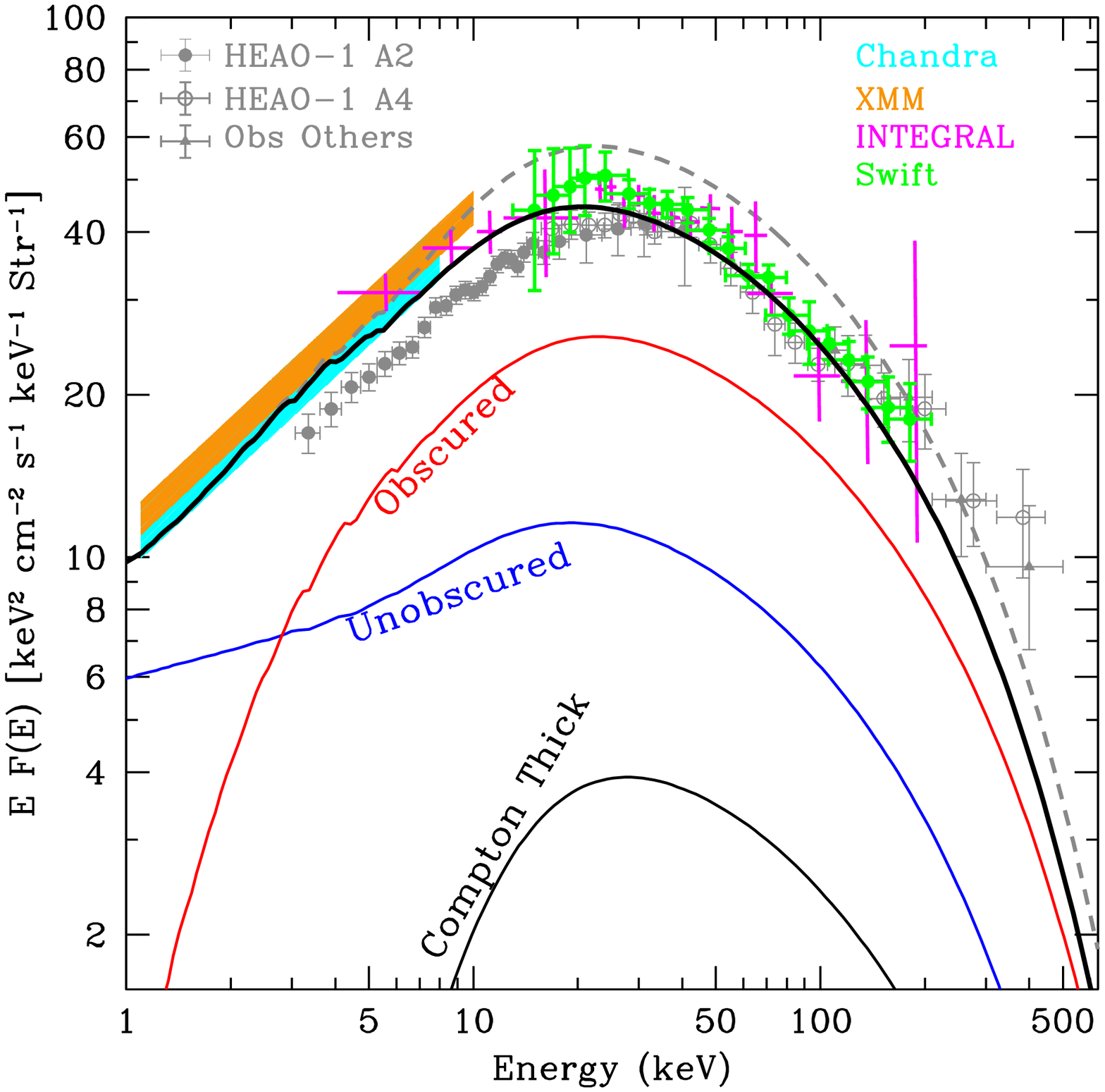}
\end{center}
\caption{Observed spectrum of the extragalactic X-ray background from 
HEAO-1 \citep{gruber99}, Chandra \citep{hickox06}, XMM \citep
{deluca04}, INTEGRAL \citep{churazov07} and Swift \citep{ajello08}
data. The {\it dashed gray line} shows the XRB spectrum from the AGN
population synthesis model of \citet{treister05b}, which assumed a 40\% higher
value for the HEAO-1 XRB normalization. The {\it thick black solid line} shows our new population
synthesis model for the XRB spectrum; the only change is the number
of CT AGN, which is reduced by a factor of 4 relative to the number in \citet{treister05b}. {\it Red}, {\it blue} and {\it
thin black} solid lines show the contribution to this model from unobscured,
obscured Compton thin and CT AGN respectively.}
\label{xrb}
\end{figure}

Since both the number density of CT AGN and the normalization of the
Compton reflection component can now be constrained independently, we
can attempt to match the observed spectrum and intensity of the
XRB. In Figure~\ref{xrb}, we show our new fit, which matches the
INTEGRAL and Swift observations at E$>$10~keV, which are $\sim$10\%
higher than the HEAO-1 normalization. The original fit of
\citet{treister05b}, which has a factor of $\sim$4 more CT AGN, is
also shown. Not surprisingly, the effects of changing the number of CT
AGN are most important in the E=10-100~keV region.

The new XRB fit matches both the INTEGRAL and Swift/BAT observations
at E$>$10~keV and the Chandra measurements at lower energies (which
are $\sim$30\% higher than the HEAO-1 A2 observations).  Recently, a
new measurement of the XRB intensity at E=1.5-7~keV using the Swift
XRT (X-ray telescope) was presented by \citet{moretti08}. These new
data confirmed that the original HEAO-1 normalization should be
increased by $\sim$30\% and $\sim$10\% at low and high energies
respectively. In contrast, the AGN population synthesis model of
\citet{gilli07} assumed the original HEAO-1 intensity at all energies,
which translates into a relatively lower contribution from unobscured
sources.  In order to produce the necessary hard spectrum,
\citet{gilli07} had to assume a relatively high number of obscured
sources at high luminosities, i.e., an unusual, inverted dependence of
the obscured fraction of AGN as a function of luminosity
\citep{hasinger08,treister08b}.

Assuming a fixed value of the Compton reflection component, how much
can the number of CT AGN change and still match the XRB, given the
existing uncertainties in the intensity measurements? The INTEGRAL
measurements of the XRB, reported by \citet{churazov07}, have
uncertainties of $\sim$5\% including both statistic and systematic
effects. Similarly, the Swift measurements have estimated errors of
$\sim$3\% \citep{ajello08}. Both measurements are consistent with each
other but are $\sim$10\% higher than the original HEAO-1
intensity. Translating this $\sim$5\% uncertainty into an uncertainty
in the number of CT AGN, we conclude that the total number of CT AGN
can be changed by a factor of 55\% and still be consistent with the
current measurements of the XRB. However, this calculation does not
include the uncertainty in the normalization of the Compton reflection
component, which is by far the dominant factor. For comparison, the
statistical errors for the measurement of the ten CT AGN detected
combining the Swift and INTEGRAL surveys correspond to an uncertainty
of $\sim$30\% \citep{gehrels86}, i.e., the direct detection of CT AGN
is much better than the XRB for determining the number of CT AGN.

\begin{figure}
\begin{center}
\plotone{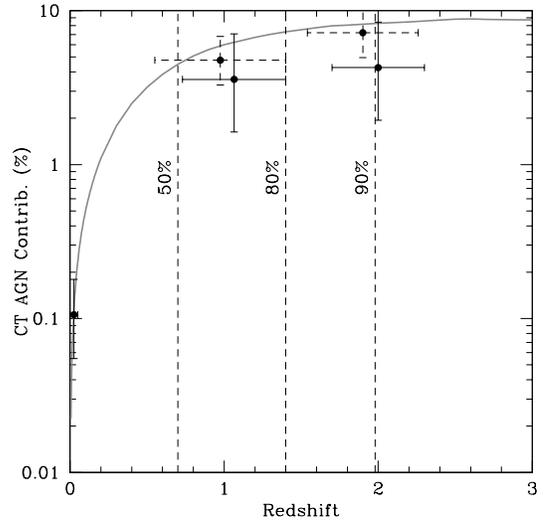}
\end{center}
\vspace{-0.5cm}
\caption{\small Cumulative fractional contribution of CT AGN to the XRB in the 14-195 keV Swift/BAT band as a function of 
redshift, from the population synthesis model presented here ({\it solid line}; see text for details). As shown by the
vertical dashed lines, 50\%, 80\% and 90\% of the total CT AGN
contribution come from sources at $z$$<$0.7, 1.4 and 2.0 respectively. Only $\sim$1\% of the total XRB intensity comes from 
CT AGN at $z$$>$2. Given the current 5\% uncertainties in the measurement of the XRB intensity, this means that the 
XRB spectrum does not constrain the number of high-redshift CT AGN at all (factor of 5 uncertainty). The data point at $z$$\sim$0 corresponds to the contribution
to the XRB by the CT AGN detected by Swift/BAT, while the data points at high redshift were obtained from the CT
AGN in the Chandra sample of \citet{tozzi06}. Solid error bars correspond to transmission-dominated
sources only, while the data points with dashed error bars include all the sources in the sample. As expected, most of the 
contribution to the XRB comes from the transmission-dominated sources, which are in general brighter. Good agreement
between our population synthesis model and observations of CT sources is found at all redshifts.}
\label{frac_ct_red}
\end{figure}

Given that the number of CT AGN in the local Universe is effectively
constrained by the Swift and INTEGRAL surveys, it is now possible to
estimate the total contribution of CT AGN to the XRB, as well as its
redshift dependence. In order to do that, we integrate the total X-ray
emission from the CT AGN in our population synthesis model and divide
it by the observed XRB intensity. To facilitate the comparison with
the local sources observed by Swift, and to make sure that the effects
of absorption are negligible, we perform this integration over the
14-195~keV band. In Figure~\ref{frac_ct_red} we show the resulting
redshift dependence of the fractional contribution to the hard XRB
radiation. As can be seen, the total contribution of CT AGN to the XRB
is $\sim$9\%, and about 50\% of it comes from sources at
$z$$<$0.7. Similarly, we conclude that $\sim$2\% of the XRB is
provided by CT AGN at $z$$>$1.4, while CT AGN at $z$$>$2 only
contribute $\la$1\% to the XRB. Conversely, the 5\% uncertainty in the
absolute measurement of the XRB intensity translates into an
uncertainty of a factor of $\sim$5 in the number of CT AGN at
$z$$>$2. Hence, the number of CT AGN at high redshift is largely
unconstrained by the XRB.

In Figure~\ref{frac_ct_red} we further compare this expected redshift
dependence to the integrated fluxes from individually-detected CT AGN.
At $z$$\simeq$0 we integrate the emission from the eight sources
detected by Swift/BAT. At higher redshifts we use the sample of CT AGN
candidates detected in X-rays in the Chandra Deep Field South reported
by \citet{tozzi06}, which includes 14 reflection-dominated AGN and six
transmission-dominated AGN with $N_H$$>$10$^{24}$~cm$^{-2}$. In the
same field, \citet{georgantopoulos07} found a total of 18 CT AGN
candidates, but only eight of them with a measured $N_H$ greater than
10$^{24}$~cm$^{-2}$; the remaining sources were selected based on
their flat X-ray spectra. All six transmission-dominated CT AGN in the
sample of \citet{tozzi06} are included in the work of
\citet{georgantopoulos07}. However, no overlap is found between the
reflection-dominated CT AGN candidates reported by \citet{tozzi06} and
the flat-spectrum sources of \citet{georgantopoulos07}. Hence, it is
possible that either selection of heavily obscured sources is highly
incomplete. In order to compare properly with the local sample, in
Figure~\ref{frac_ct_red} we show separately the contribution from the
transmission-dominated AGN and from all sources in the sample of
\citet{tozzi06}. We separated the sample at $z$=1.5, to have the same
number of sources in each redshift bin. As expected, most of the
contribution to the XRB comes from the transmission-dominated sources,
which are in general brighter in the X-ray band. The agreement at low
redshift is not surprising, since by construction our model was
adjusted to match the Swift/BAT observations. However, it is very
interesting that also for the high redshift sources the calculated
contribution of CT AGN to the XRB agrees well with the limits from
deep surveys.

\section{High-redshift CT AGN}

As shown in the previous section, the number of CT AGN at high
redshift is largely unconstrained by the XRB or by current hard X-ray
surveys. Since a large fraction of the absorbed energy in heavily
obscured AGN is re-emitted at mid-IR wavelengths, deep Spitzer data
have been used to find CT AGN candidates at $z$$\geq$2. \citet{daddi07} used the
excess in mid-IR luminosity (compared to UV estimates of star
formation rates) to find obscured AGN not individually detected in
X-rays. Similarly, \citet{fiore08} used a combination of red
optical-to-near-IR colors and high 24~$\mu$m luminosity to select CT
AGN candidates at $z$$\sim$2. In both cases, very high source
densities have been estimated for mid-IR-selected CT AGN, e.g.,
\citet{daddi07} reported a sky density of $\sim$3,200~deg$^{-2}$,
similar to that of all previously known AGN at those redshifts in the
Chandra deep fields. Somewhat surprisingly, very little overlap is
found between the two selection methods, implying the possible
presence of interlopers.  If these candidates are confirmed, a very
large number of CT AGN exist at high redshift, considerably larger
than the local population. This is qualitatively consistent with the
evolution in obscuration detected by \citet{treister06b} for
Compton-thin sources. Recently, \citet{alexander08} reported the
confirmation of seven CT AGN using optical and mid-IR spectroscopy in
the Chandra Deep Field North region, implying a similarly high density
for CT AGN at high redshift.

\begin{figure}
\begin{center}
\plotone{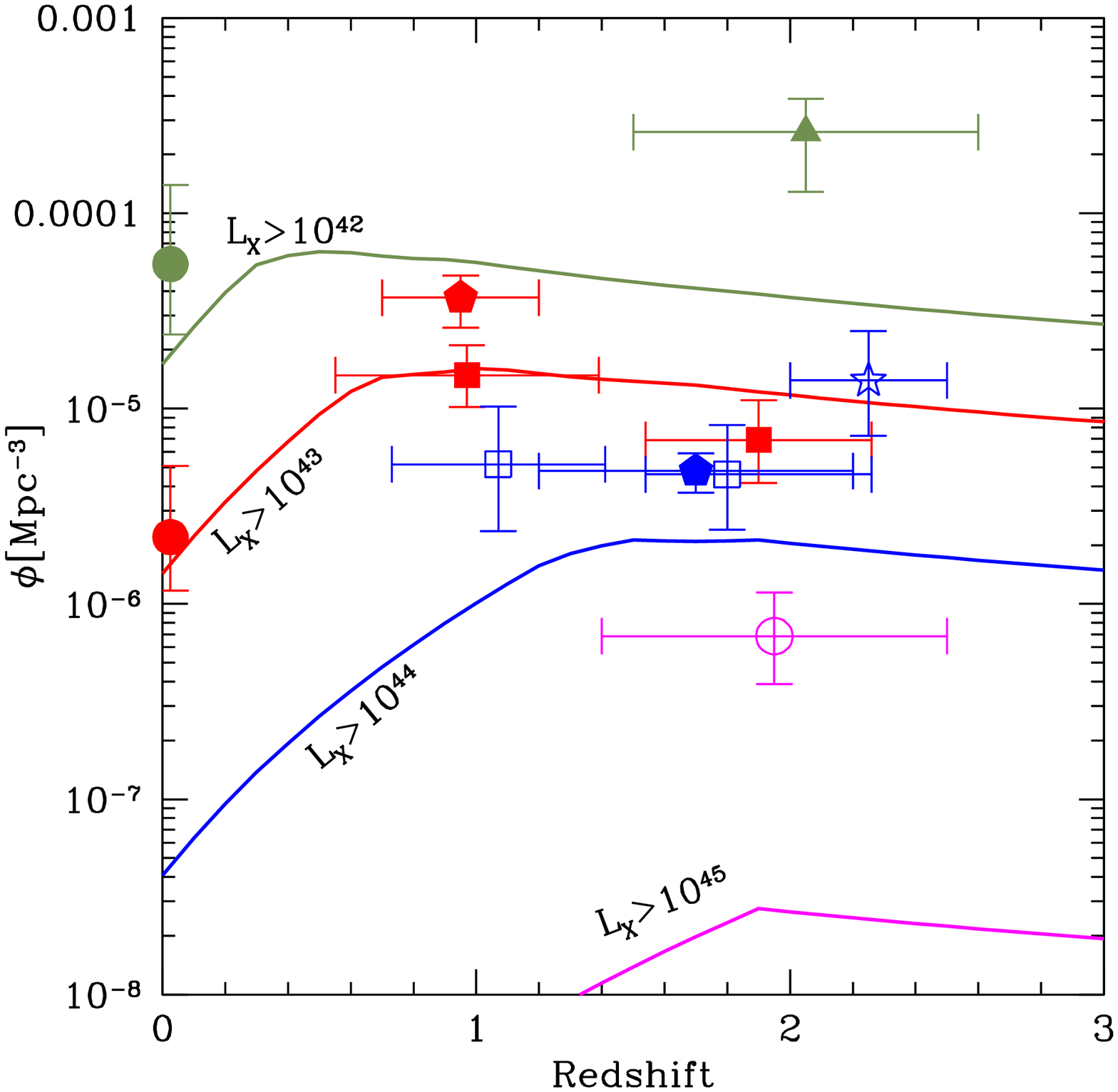}
\end{center}
\caption{Comoving density of CT AGN as a function of redshift in several luminosity bins.
Measured values (details in text)
are shown by: {\it open circle:}
\citet{polletta06}, {\it squares:} \citet{tozzi06}, {\it filled triangle:} \citet{daddi07}, {\it pentagons:}
\citet{fiore09}, {\it star:} \citet{alexander08}, {\it filled circle:} Swift/BAT and INTEGRAL, 
for luminosity limits indicated by color: {\it magenta}: $L_X$$>$10$^{45}$~erg~s$^{-1}$, 
{\it blue}: $L_X$$>$10$^{44}$~erg~s$^{-1}$, {\it red}: $L_X$$>$10$^{43}$~erg~s$^{-1}$
and {\it green}: $L_X$$>$10$^{42}$~erg~s$^{-1}$. {\it Solid lines} show the space densities at corresponding luminosities for 
the \citet{treister05b} model including the evolution of obscured AGN reported 
by \citet{treister06b}, with the number of CT AGN adjusted to the observed local value (present paper). 
At high redshifts, 
a relatively large density of CT AGN is observed, compared to 
expectations from the evolving luminosity function. This suggests
that the local sample is incomplete; or that the high-redshift 
infrared-selected samples include a large number of interlopers; 
or that CT AGN follow a different evolution than Compton-thin
sources. }
\label{com_dens}
\end{figure}

In order to quantify the density of CT AGN at high redshift, and to
compare with local observations and AGN luminosity functions, in
Figure~\ref{com_dens} we present the available measurements of the
comoving volume density of CT AGN candidates as a function of
redshift. Our measurement of the density of CT AGN at $z$=0 is
$\sim$2.2$\times$10$^{-6}$ Mpc$^{-3}$, for sources with
$L_X>$10$^{43}$~erg~s$^{-1}$, as shown in \S 2.1. At higher redshifts
and luminosities, we infer the space density from several
samples. Five X-ray-selected CT AGN candidates were found in the
Spitzer Wide-area InfraRed Extragalactic survey (SWIRE;
\citealp{polletta06}); these are transmission-dominated CT AGN and all
of them have intrinsic X-ray luminosities greater than
10$^{45}$~erg~s$^{-1}$ (magenta open circle, Fig.~\ref{com_dens}).
Twenty X-ray-selected CT AGN were found by \citet{tozzi06} in the
Chandra Deep Field South; in this case, we separated the sample into
two redshift bins and computed the comoving number density separately
for sources with $L_X$$>$10$^{43}$ and $L_X$$>$10$^{44}$~erg~s$^{-1}$
(squares, Fig.~\ref{com_dens}).  Finally, we also show the comoving
number density estimated from the mid-IR CT AGN candidates in the
samples of \citet{daddi07}, \citet{fiore09} and \citet{alexander08}.

The expected comoving number density for CT AGN as a function of
redshift was computed using the models of
\citet{treister05b}. Briefly, we used the hard (2-10~keV) X-ray
luminosity function and AGN evolution of \citet{ueda03}, with the
$N_H$ distribution and luminosity dependence of the fraction of
obscured AGN of \citet{treister05b}. In addition, we include the
evolution of the relative number of obscured sources reported by
\citet{treister06b}.  The normalization of the relative number of CT
AGN was chosen to match the observed numbers at $z$$\sim$0. This model
fits the observed XRB spectral shape and normalization, as shown in
Figure~\ref{xrb}. The resulting comoving density of CT AGN as a
function of redshift is shown in Figure~\ref{com_dens} for sources
with $L_X$$>$10$^{42}$, 10$^{43}$, and 10$^{44}$~erg~s$^{-1}$.

While the observed density of X-ray-selected CT AGN with
$L_X$$>$10$^{43}$~erg~s$^{-1}$ is in pretty good agreement with the
expectations at all redshifts, at higher luminosities the observed
values are mostly higher than the expectations. In fact, the
comoving density for $L_X$$>$10$^{45}$~erg~s$^{-1}$ sources from SWIRE
is almost two orders of magnitude higher than the expected
value. However, it is important to note that this comoving density is
derived from CT AGN candidates based on the spectral properties
derived from low signal-to-noise observations, and also the number of
sources detected is small, so the uncertainties are large. The values
for X-ray-selected CT AGN with $L_X$$>$10$^{44}$~erg~s$^{-1}$ are also
higher than expectations, but in this case only by factors of
$\sim$2-3. Similarly, the densities inferred from mid-IR-selected CT
AGN are systematically higher than the expected values at all
luminosities, typically by one order of magnitude.

This discrepancy between expectations and observations at high
redshift can be interpreted in several ways. One obvious possibility
is that the observations at low redshift are missing a significant
number of CT AGN, which are included in the high-redshift samples. In
fact, we have shown before that the high-energy surveys performed by
INTEGRAL and Swift are mostly complete for transmission-dominated
sources, but miss a significant fraction of the reflection-dominated
AGN. While the sample of \citet{risaliti99} includes these sources, it
is based on pointed observations and hence it is highly incomplete as
well (Fig.~\ref{logn_s}). Also, it is very likely that the
high-redshift IR-selected samples include both transmission and
reflection-dominated sources. Another possibility is that the
high-redshift samples, both X-ray- and mid-IR-selected, include a
significant number of interlopers. These could be either less obscured
AGN in the case of X-ray selection, or non-active galaxies undergoing
significant but dusty star formation, and thus showing high mid-IR
luminosities not due to AGN activity. Finally, it is possible that CT
AGN follow a different evolution than Compton-thin sources. It is
important to note that the (1+$z$)$^{0.4}$ evolution in the ratio of
obscured to unobscured AGN found by \citet{treister06b} for
Compton-thin sources is already included in the predicted volume
densities. Hence, if both the low and high redshift observed samples
are not systematically missing a significant number of sources, the
excess of CT AGN at high redshift implies a different (stronger)
evolution for these heavily obscured sources than for obscured but
Compton-thin AGN.

\section{The Density of Supermassive Black Holes}

Because AGN are powered by accretion of gas onto a supermassive black
hole, the AGN luminosity function represents the history of cosmic
accretion \citep{soltan82}. Hence, the AGN bolometric luminosity can
be converted into a mass accretion rate, assuming an efficiency for
the conversion $\epsilon$=L/$\dot{m}$c$^2$ (typically,
$\epsilon$$\simeq$0.1). Then, the comoving black hole mass density can
be written (following equation 17 of \citealp{yu02}) as:

\begin{small}

\begin{equation}
\rho (z)=\int_z^{\infty}\frac{dt}{dz}dz\int_{L_{min}}^{L_{max}}\frac{(1-\epsilon)\textnormal{BC}(L_X)L_X}{\epsilon c^2}\Psi(L,z)\int_{N_{H,min}}^{N_{H,max}}f(N_H,L)dN_HdL,
\end{equation}

\end{small}

\noindent where $\Psi (L,z)$ is the evolving AGN luminosity function,
BC$(L_X)$ is the bolometric correction starting from the 2-10~keV
luminosity, and $f(N_H,L)$ is the ``$N_H$ function,'' or the fraction
of sources at a given luminosity with a given $N_H$. For this
calculation we used the 2-10~keV luminosity function of \citet{ueda03}
and the $N_H$ function with a luminosity dependence described in
Section 3.2 of \citet{treister05b}. The bolometric correction was
calculated using the spectral energy distribution of a completely
unobscured AGN as specified by \citet{treister06a}, as appropriate for
the unified model of AGN. Additionally, we updated the spectral
library with the new relation between X-ray (at 2~keV) and UV
(2500\AA~) luminosities using the value of the slope of the power-law
extrapolation reported by \citet{steffen06},

\begin{equation}
\alpha_{ox}=(-0.077\pm 0.015)\log(L_{\textnormal{2 keV}})+(0.492\pm 0.387).
\end{equation}

\noindent With these assumptions, the bolometric correction ranges from $\sim$25 for
$L_X$=10$^{42}$~ergs~s$^{-1}$ to $\sim$100 for
$L_X$=10$^{45}$~ergs~s$^{-1}$, in approximate (factor of $\sim$2)
agreement with the values assumed by other authors (e.g.,
\citealp{kuraszkiewicz03,marconi04,barger05,hopkins07}).

In previous works (e.g., \citealp{yu02}), the AGN luminosity function
was integrated from $L_{min}$=0 to $L_{max}$=$\infty$, which leads to
very large extrapolations in particular at the faint end. In the
present case, we use the same integration limits used by
\citet{treister05b} in their AGN population synthesis model, namely,
$L_{min}$=10$^{41.5}$~ergs~s$^{-1}$,
$L_{max}$=10$^{48}$~ergs~s$^{-1}$, $N_{H,min}$=10$^{20}$~cm$^{-2}$ and
$N_{H,max}$=10$^{25}$~cm$^{-2}$. The number of CT AGN in this model is
matched to the INTEGRAL and Swift results, as reported above. With
these assumptions, and using the typical value of $\epsilon$=0.1, we
obtain a value for the local black hole mass density of
$\rho$(z=0)=4.5$\times$ 10$^5$ M$_\odot$~Mpc$^{-3}$. This calculation
agrees well with the values estimated from observations:
$\rho$=4.6$^{+1.9}_{-1.4}$$\times$10$^5$~M$_\odot$Mpc$^{-3}$
\citep{marconi04} and
$\rho$=(3.2-5.4)$\times$10$^5$~M$_\odot$Mpc$^{-3}$ \citep{shankar08a}.

\begin{figure}
\begin{center}
\plotone{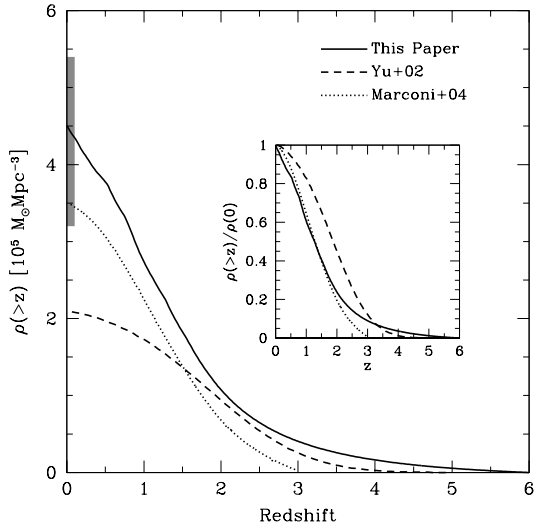}
\end{center}
\caption{Black hole mass density as a function of redshift (inset 
shows the same curves normalized to their values at $z$=0),
assuming an efficiency $\epsilon$$\equiv$$L/\dot{m} c^2$$=0.1$.
The {\it solid lines} show the evolution for the population synthesis model described in this 
paper, while the {\it dotted lines} show a similar calculation presented by
\citet{marconi04}. In both cases the AGN luminosity function of \citet{ueda03}
was used, and the only differences are the number of CT AGN (4 times more 
in the model of \citealp{marconi04}) and the assumed bolometric correction
($\sim3$ times higher for our calculation). The
{\it gray rectangle} at $z$=0 shows the range of values consistent with observations,
as reported by \citet{shankar08a}. For comparison, the {\it dashed lines} show
the black hole mass density estimated by \citet{yu02} which considered 
only high-luminosity
unobscured sources.}
\label{bhmass_w_z}
\end{figure}

In Figure~\ref{bhmass_w_z} we present the black hole mass density as a
function of redshift estimated from our calculation, together with the
curves presented by \citet{marconi04} and \citet{yu02}. The main
differences with the work of \citet{marconi04} are in the number of CT
AGN (they assumed 4 times more), the assumed bolometric correction
(our is $\sim$3 times higher at low luminosities) and the redshift
limit of the integration. Of these, the bolometric correction
dominates, such that our derived local black holes mass density is
slightly larger, even with the reduced number of CT AGN. Note that the
bolometric correction of \citet{marconi04} was obtained from
observations of high-luminosity sources only, while our bolometric
correction was tested by observations of fainter sources as well
\citep{treister06a}. A remarkably good agreement is found between our
results and the recent work of \citet{shankar08a}.  Compared to
\citet{yu02}, we find twice the local integrated black hole mass
density, because they used an optical quasar luminosity function and
evolution, which peaks at a higher redshift, $z\sim$2, and evolves
strongly. In contrast, the \citet{ueda03} luminosity function, which
includes lower luminosity and obscured sources, peaks at
$z\simeq$1.1. In our calculation, the vast majority of the black hole
growth occurs at low redshift ($\sim$50\% from $z=1.3$ to 0), which
matches observations of AGN detected in X-rays (e.g,
\citealp{barger01}).

The space density of CT AGN is consistent with, but cannot be
constrained by, the observed local black hole mass density. In
addition, analogous to the weakness of the XRB integral constraint on
the number of CT AGN, numerous degeneracies with other parameters,
like the assumed bolometric correction and efficiency, are
important. Even taking into account only the uncertainties in the
local black hole mass density, we could still increase the number of
CT AGN in the local Universe by factors of $\sim$3. Hence, we can
conclude that direct observations of CT AGN at high energies, like the
INTEGRAL and Swift observations discussed here, are currently the only
way to constrain the population of heavily obscured supermassive black
holes.

\section{Conclusions}

In this paper we constrain the space density of CT AGN in the local
Universe using the recently-available wide-area surveys at high
energies performed by INTEGRAL/IBIS and Swift/BAT. A total of ten CT
AGN at $z$$<$0.03 were found by either INTEGRAL and/or Swift. These
observations are complete for transmission-dominated CT AGN, but are
probably still missing heavily obscured sources with
$N_H$$>$10$^{25}$~cm$^{-2}$.  We find that the space density of local
CT AGN follow a Euclidean distribution with a normalization of
$\sim$10$^{-4}$~deg$^{-2}$ at fluxes of $\sim$5$\times$10$^{-11}$ and
$\sim$9$\times$10$^{-11}$~erg~cm$^{-2}$s$^{-1}$ in the 17-60~keV and
14-195~keV bands, respectively. This is about 3-4 times smaller than
the values expected from recent AGN population synthesis models that
fit the extragalactic XRB \citep{treister05b,gilli07}, and thus
modifications to those models is required.

We present here a new population synthesis model for the XRB, with the
number of CT AGN constrained by the INTEGRAL and Swift number
counts. We find that the fraction of AGN that are Compton thick at
$F_{20-100 keV}$$\sim$10$^{-11}$~erg~cm$^{-2}$s$^{-1}$ is $\sim$5\%.
We show that the XRB by itself cannot be used to constrain the number
of CT AGN, mainly due to degeneracies with other parameters, the most
important of which is the normalization of the Compton reflection
component.  We find that the total contribution of CT AGN to the XRB
is $\sim$9\%, with only $\lesssim$1\% from CT AGN at $z$$>$2. Hence,
taking into account the 5\% uncertainty in the XRB intensity
measurements, the number of CT AGN at high redshift is essentially
unconstrained by the XRB, even if all the other parameters could be
fixed.

We calculate the local black hole mass density inferred from AGN
activity using Soltan's argument \citep{soltan82}, taking into account
the contribution from CT AGN estimated in this work.  For an accretion
efficiency $\epsilon$$\equiv$$L/\dot{m} c^2$$=0.1$, we find an integrated
local black hole mass density of 4.5$\times$ 10$^5$
M$_\odot$~Mpc$^{-3}$, in excellent agreement with recent estimates
based on measured masses of local dormant black holes.  Considering
the current uncertainties in these estimates, we conclude that only
the direct observations of CT AGN such as those discussed in this
paper can effectively constrain the number of heavily obscured AGN.
Based on the number density of CT AGN presented here, our best
estimate of the fractional contribution of CT AGN to the total
accreted black hole mass is $<10$\%.

Using a combination of X-ray and mid-IR selection, the space density
of CT AGN at high redshift is starting to be constrained. We find that
observed densities are systematically higher than expected from the
evolving AGN luminosity function measured from less obscured sources,
assuming $N_H$-independent evolution of the local CT AGN population.
This can be explained in three ways, any or all of which could be the
case.  First, the local sample might be incomplete, particularly
because even hard X-ray selection is biased against
reflection-dominated CT AGN.  Second, the high-redshift samples may be
contaminated by strongly star-forming galaxies or other interlopers.
Third, CT AGN may evolve more strongly than less-absorbed sources,
implying a relatively larger number of CT AGN in the early
Universe. To decide this question requires the help of observations
with the new generation space-based hard X-ray observatories. While
mid-IR selection of heavily obscured AGN is very promising, these
samples are inevitably affected by the presence of interlopers, in
particular from star-forming galaxies, and by the lack of accurate
measurements of the amount of obscuration.

Hard X-ray selection provides a cleaner sample of CT AGN, since $N_H$
values can be measured directly and there is almost no contamination
from star-forming galaxies at these energies. Several different
approaches are currently being planned for the next generation of
high-energy (E$>$10~keV) missions, to provide a large and complete
sample of CT AGN up to $z$$\sim$3. The Energetic X-ray Imaging Survey
Telescope\footnote{More information about EXIST can be found at
http://exist.gsfc.nasa.gov/} (EXIST; \citealp{grindlay05}) will
perform an all-sky survey in the 20-80~keV energy band to flux limits
of $\sim$6$\times$10$^{-13}$~erg~cm$^{-2}$s$^{-1}$, finding thousands
of heavily obscured AGN up to $z$$\sim$1 and high-luminosity CT
quasars at all redshifts. With a complementary approach, the Nuclear
Spectroscopic Telescope Array\footnote{The NuSTAR website URL is
http://www.nustar.caltech.edu/} (NuSTAR; \citealp{harrison08}), with a
scheduled launch date of August 2011, will perform targeted
observations of fields of $\sim$1~deg$^2$ to flux limits of
$\sim$2$\times$10$^{-14}$~erg~cm$^{-2}$s$^{-1}$, hence $\sim$20 times
deeper than EXIST, in the 6-79 keV band, for exposure times of $\sim$1
Msec; these observations will be able to find low-luminosity CT AGN up
to $z$$\sim$2-3. Similarly, the planned New X-ray
Telescope\footnote{http://www.astro.isas.ac.jp/future/NeXT/} (NeXT;
\citealp{takahashi08}), scheduled for launch in 2013 will provide
imaging and spectroscopy in the 5-80 keV energy band with an angular
resolution $<$1.7$'$ and a spectral resolution of $\sim$1.5
keV. Another focusing hard X-ray observatory, Simbol-X\footnote{More
details can be found at http://smsc.cnes.fr/SIMBOLX/}, is targeted for
launch in 2014 \citep{ferrando04}. Simbol-X will perform pointed
observations with a field of view of $\sim$12$'$ and an angular
resolution of $\sim$30''.

Finally, it is important to note that for $z$$\sim$2 the Chandra and
XMM observed energy band of 2-10~keV corresponds to a rest-frame
energy of $\sim$6-30~keV, so the effects of obscuration are less
important. Unfortunately even the deepest Chandra data available now
only detect a few photons for the CT AGN candidates at $z$$\sim$2
(e.g., \citealp{tozzi06}), thus preventing detailed spectral fitting
that could provide a deeper physical understanding of the nature of
these sources. The proposed International X-ray
Observatory\footnote{http://ixo.gsfc.nasa.gov/} (IXO) will provide an
outstanding opportunity to study these highly-obscured high-redshift
sources. As reported by \citet{alexander08}, the IXO will be able to
detect thousands of photons for the CT AGN detected in the Chandra
Deep Fields observations for similar, $\sim$1 Msec, exposure times,
yielding high signal-to-noise spectra for these sources. Deep
observations at high energies with NuSTAR, EXIST and Simbol-X will
provide large samples of heavily obscured AGN at $z$$\sim$1--3, while
the improved sensitivity and spectral resolution of the IXO will allow
us to study in detail the spectra of CT AGN at $z$$\sim$2--3.

\acknowledgments

We thank the anonymous referee for very useful comments. We acknowledge support from NASA/INTEGRAL grants NNG05GM79G
and NNX08AE15G. Support for the work of ET was provided by the National Aeronautics and
Space Administration through Chandra Postdoctoral Fellowship Award
Number PF8-90055 issued by the Chandra X-ray Observatory Center, which
is operated by the Smithsonian Astrophysical Observatory for and on
behalf of the National Aeronautics Space Administration under contract
NAS8-03060. SV acknowledges support from a graduate research
scholarship awarded by the Natural Science and Engineering Research
Council of Canada (NSERC).






\end{document}